\documentclass[journal]{IEEEtran}

\usepackage{amsmath}
\usepackage{graphicx}
\usepackage{textcomp}
\usepackage{cite}
\usepackage{amssymb}
\usepackage{multirow}
\usepackage{color}
\usepackage{epstopdf}

\ifCLASSINFOpdf

\else

\fi

\begin{document}

\title{Model-Free MLE Estimation for Online Rotor Angle Stability Assessment with PMU Data}

\author{Shaopan Wei,~\IEEEmembership{Student Member,~IEEE,}
	    Ming Yang,~\IEEEmembership{Member,~IEEE,}
	    Junjian Qi,~\IEEEmembership{Member,~IEEE,}
	    Jianhui~Wang,~\IEEEmembership{Senior Member,~IEEE,}
        Shiying Ma,
        and Xueshan Han
		\thanks{This work was supported by the National Basic Research Program of China (973 Program) under Grant 2013CB228205, State Grid Corporation of China under Grant XT71-15-056, and the National Science Foundation of China under Grant 51007047 and 51477091.}		\thanks{M. Yang (Corresponding Author) is with Key Laboratory of Power System Intelligent Dispatch and Control, Shandong University, Jinan, Shandong 250061 China. He was a visiting scholar with Argonne National Laboratory, Argonne, IL 60439 USA (e-mail: myang@sdu.edu.cn).}
		\thanks{S. Wei and X. Han are with Key Laboratory of Power System Intelligent Dispatch and Control, Shandong University, Jinan, Shandong 250061 China (e-mail: spw\_sdu@sina.com; xshan@sdu.edu.cn).} 
		\thanks{J. Qi and J. Wang are with the Energy Systems Division, Argonne National Laboratory, Argonne, IL 60439 USA (e-mail: jqi@anl.gov; jianhui.wang@anl.gov).}
		\thanks{S. Ma is with Institute of Electric Power System, China Electric Power Research Institute, Haidian District, Beijing 100192 China (e-mail: mashiy@epri.sgcc.com.cn).}}
\maketitle

\begin{abstract}
Recent research has demonstrated that the rotor angle stability can be assessed by identifying the sign of the system's maximal Lyapunov exponent (MLE). 
A positive (negative) MLE implies unstable (stable) rotor angle dynamics.
However, because the MLE may fluctuate between positive and negative values for a long time after a severe disturbance, 
it is difficult to determine the system stability 
when observing a positive or negative MLE without knowing its further fluctuation trend.
In this paper, a new approach for online rotor angle stability assessment is proposed to address this problem. 
The MLE is estimated by a recursive least square (RLS) based method based on real-time rotor angle measurements, and 
two critical parameters, the Theiler window and the MLE estimation initial time step, 
are carefully chosen to make sure the calculated MLE curves present distinct features for different stability conditions. 
By using the proposed stability assessment criteria, the developed approach can 
provide timely and reliable assessment of the rotor angle stability.
Extensive tests on the New-England 39-bus system and the Northeast Power Coordinating Council 140-bus system verify the effectiveness of the proposed approach.
\end{abstract}


\IEEEpeerreviewmaketitle

\section{Introduction}

\IEEEPARstart{T}{ransient} rotor angle stability refers to the ability of synchronous generators of an interconnected power system to remain in synchronism after a severe disturbance \cite{kundur2004definition}.
With the development of 
synchrophasor technologies, utilities are 
now able to track rotor angle deviations and take actions to respond to emergency events.
However, since the dynamics of power systems are complex, 
online rotor angle stability assessment is still very challenging \cite{yan2011pmu,dasgupta2015pmu}.

In \cite{centeno1997adaptive}, an adaptive out-of-step relay is proposed for the Florida-Georgia system. 
The equal area criterion is applied to change the settings of the protection system based on phasor measurement unit (PMU) measurements.
In \cite{chow2007synchronized}, the dynamics of the power transfer paths are monitored based on the energy functions of the two-machine equivalent system, 
and the PMU data are used to identify the parameters of the energy functions.
In \cite{liu2000new}, PMU measurements are used as inputs for estimating the differential/algebraic equation model to predict the post-fault dynamics.
In \cite{sun2007online}, an online dynamic security assessment scheme is proposed based on self-adaptive decision trees, where the PMU data are used for online identification of the system critical attributes.
In \cite{del2007estimation}, the rotor angle stability is estimated by using artificial neural networks and 
the measured voltage and current phasors are used as inputs of the offline trained estimation model.
In \cite{kamwa2009development}, a systematic scheme for building fuzzy rule-based classifiers for fast stability assessment is proposed.
By testing on a large and highly diversified database, it is demonstrated that the analysis of post-fault short-term PMU data 
can extract useful features satisfying the requirements of stability assessment.

Lyapunov exponents (LEs) those characterize the separation rate of infinitesimally close trajectories are important indices for quantifying the stability of dynamical systems.
If the system's maximal Lyapunov exponent (MLE) is positive, the system is unstable, and vice versa.
LEs are first applied to power system stability analysis in \cite{liu1994detection}, in which 
it is verified that LEs can predict the out-of-step conditions of power systems. 
In \cite{yan2011pmu}, a model-based MLE method is proposed 
for online prediction of the rotor angle stability with PMU measurements.
The work builds solid analytical foundations for the LE-based rotor angle stability assessment.
In \cite{khaitan2013vantage}, the LEs are calculated with dynamic component and network models to identify the coherent groups of generators.

Although the model-based MLE estimation approaches have made significant progresses on online rotor angle stability assessment, 
they are usually computationally expensive especially when applied to large power systems.
Therefore, two model-free MLE estimation approaches have been proposed for transient voltage stability assessment \cite{dasgupta2013real} and rotor angle stability assessment \cite{dasgupta2015pmu}, 
for which the MLEs can be estimated by only using PMU measurements.

The model-free LE-based stability assessment approaches are attractive, because they can eliminate model errors and simplify the calculation.
However, when applying these approaches, a time window has to be pre-specified for the MLE observation.
The window size is crucial for obtaining reliable and timely assessment results, i.e., 
too small window size will lead to unreliable assessment results while too large window size will lead to accurate but untimely results.
The window size is difficult to be determined in advance, 
because the estimated MLEs may fluctuate between positive and negative values for quite a long time after disturbances and the window size should change with different fault scenarios.

In this paper, a LE-based model-free rotor angle stability assessment approach is proposed.
The MLEs are estimated by a \textcolor{black}{recursive least square} (RLS) based method based on real-time rotor angle measurements.
By properly choosing two critical parameters according to the characteristics of the relative rotor angles of the selected generator pairs, 
the calculated MLE curves will present distinct features for different stability conditions, 
based on which the stability criteria are correspondingly designed to capture the MLE features and perform online rotor angle stability assessment.
Compared with the existing approaches, the proposed approach does not need a pre-specified time window to identify the sign of MLEs.
Instead, the proposed approach can always make a reliable and timely assessment as soon as the crucial features are observed from the estimated MLE curves.

The remainder of this paper is organized as follows. 
Section II introduces 
the theoretical basis of model-free MLE estimation.
Section III proposes 
a rotor angle stability assessment approach, and discusses the parameter selection principles and stability criteria.
In Section IV, simulation results on the New-England 39-bus system and the Northeast Power Coordinating Council (NPCC) 140-bus system are presented to validate the effectiveness of the proposed approach.
Finally, the conclusions are drawn in Section V.

\section{MLE Estimation From Time Series}

LEs 
can reflect the exponential divergence or convergence of neighboring trajectories in the state space of a dynamic system \cite{skokos2016chaos}.
An $N$-dimensional dynamic system has $N$ LEs and the largest one is defined as the MLE of the system.
MLE is a useful indicator of system stability:
A positive MLE indicates unstable system dynamics while a negative MLE indicates asymptotically stable dynamics \cite{yan2011pmu}.
The MLE can be estimated by using Jacobian matrix based (model-based) methods \cite{skokos2016chaos,yan2011pmu} or direct model-free methods \cite{skokos2016chaos,dasgupta2013real,dasgupta2015pmu}.
Compared with the Jacobin matrix based methods, direct methods 
are more suitable for online stability assessment mainly because they do not need 
the repeated computing of the Jacobian matrix or even
the dynamic model of the system.

\begin{figure}[!htb]
	\centering
	\includegraphics[width=1.8in]{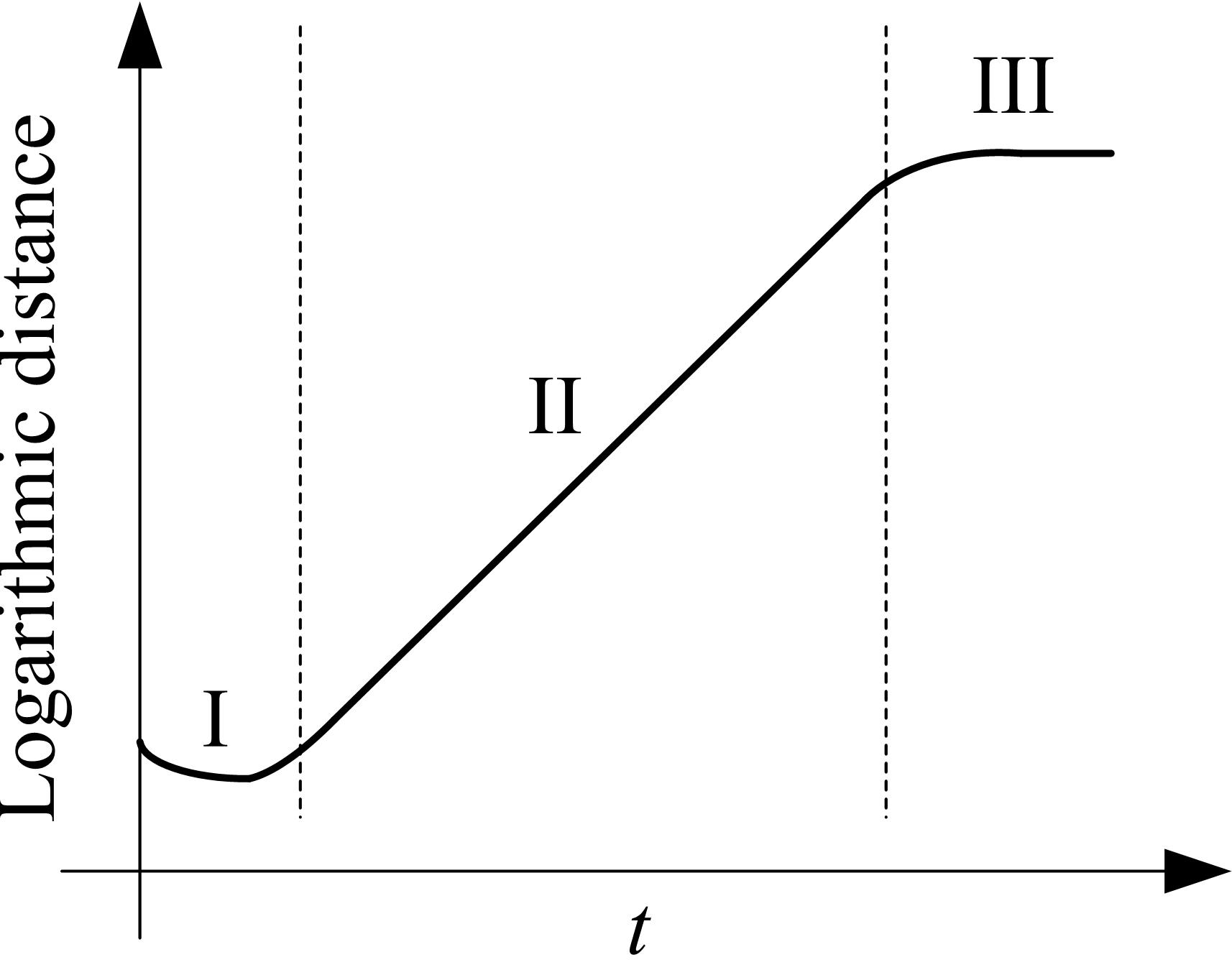}
	\caption{Logarithmic distance of neighboring states on different trajectories.}
	\label{F1}
\end{figure}

According to Oseledec's multiplicative ergodic theorem \cite{skokos2016chaos, oseledec1968multiplicative,rosenstein1993practical}, 
for a reference point $\boldsymbol{X}_0$ and its neighboring point $\boldsymbol{X}_{m(0)}$ chosen from the state space of a nonlinear dynamic system, the distance between the trajectories emerged from $\boldsymbol{X}_0$ and $\boldsymbol{X}_{m(0)}$, i.e., the original trajectory and the neighboring trajectory, will have three different growth phases as illustrated in Fig. \ref{F1}.
In Phase I, the difference vector between the states of the trajectories gradually converges towards the most expanding direction, and the distance between the trajectories will exhibit fluctuations. 
In Phase II, the distance experiences an 
exponential growth characterized by the MLE, which corresponds to 
a linear segment in the semi-logarithmic plot.
Finally, in Phase III the separation of the trajectories is saturated and the distance converges to a constant value.

As proved by \cite{skokos2016chaos} and \cite{gao1993local,kantz1994robust,rosenstein1993practical}, the MLE can be estimated from the mean logarithmic separation rate of the trajectories in Phase II as
\begin{align}\label{eq:1}
&\lambda_k \thickapprox \frac{1}{k\Delta t} \log\left(\frac{d\left(m(n),n,k\right)}{d\left(m(n),n,0\right)}\right) \nonumber
\\
& \ \ \ = \frac{1}{k\Delta t} \log\left(\frac{||\boldsymbol{X}_{m(n)+k}-\boldsymbol{X}_{n+k}||}{||\boldsymbol{X}_{m(n)}-\boldsymbol{X}_{n}||}\right) \nonumber,
\\
&\boldsymbol{X}_{n}, \boldsymbol{X}_{m(n)}, \boldsymbol{X}_{n+k}, \textrm{and}\, \boldsymbol{X}_{m(n)+k} \in\text{Phase II},
\end{align} 
where
$\lambda_k$ is the estimated MLE, 
$k$ is the lagged time steps for the MLE estimation,
$\Delta t$ is the time duration for each time step,
$\boldsymbol{X}_{n}$ and $\boldsymbol{X}_{m(n)}$ are the initial points for the MLE estimation on the original and neighboring trajectories, respectively,
$\boldsymbol{X}_{n+k}$ is the $k$th point behind $\boldsymbol{X}_{n}$ on the original trajectory,
$\boldsymbol{X}_{m(n)+k}$ is the $k$th point behind $\boldsymbol{X}_{m(n)}$ on the neighboring trajectory,
$d\left(m(n),n,0\right)$ is the Euclidean distance between the MLE estimation initial points,
$d\left(m(n),n,k\right)$ is the Euclidean distance between the $k$th points behind the MLE estimation initial points, 
and $||A-B||$ denotes the Euclidean distance between points $A$ and $B$.


\begin{figure}[!b]
	\centering
	\includegraphics[width=3.3in]{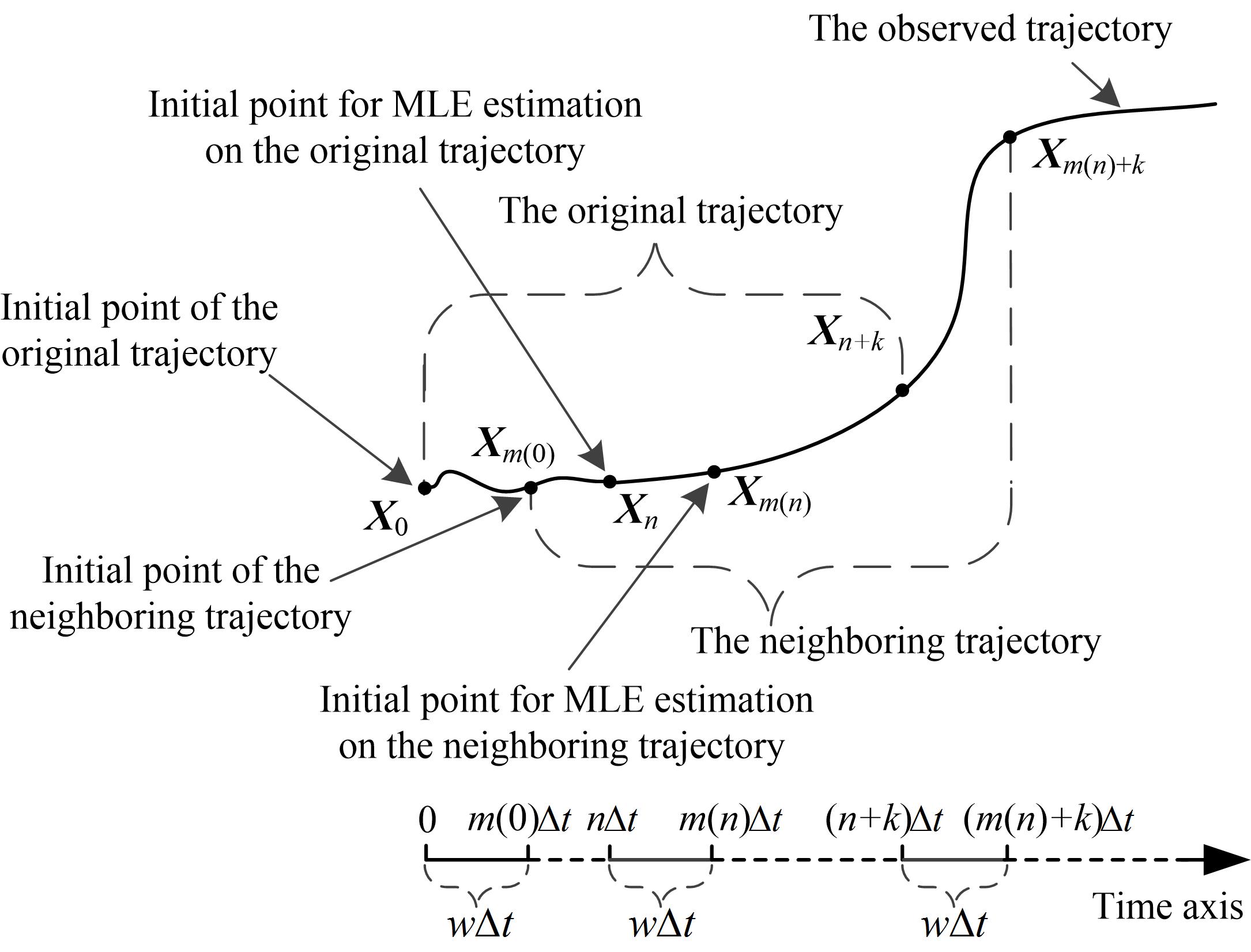}
	\caption{Trajectories and the effect of the Theiler window.}
	\label{F2}
\end{figure}

It should be noted that the original and the neighboring trajectories are usually 
from the same observed time series with different initial points, as illustrated in Fig. \ref{F2}. 
In order to make sure the trajectories are temporally separated and thus can be seen as different trajectories, 
the trajectory initial points should satisfy $|m(0)-0|>w$, where $w$ is called the Theiler window \cite{skokos2016chaos} 
and should be determined according to the characteristics of the system.

\section{LE-Based Rotor Angle Stability Assessment}

Here we propose a rotor angle stability assessment approach based on MLE estimation. 
It is data-driven 
and can perform online stability assessment only 
using the rotor angle and rotor speed of the generators 
\cite{yan2011pmu,dasgupta2015pmu}.
Although the rotor angle and rotor speed may not be directly available from PMU measurements, 
they can be estimated by various dynamic state estimation methods \cite{huang2007feasibility,ghahremani2011dynamic,sun2016power,ghahremani2016local,qi2016}.

\subsection{State Variable Selection}

Even a moderate-size power system may still have hundreds of state variables. 
Due to both the calculation intractability and 
the insufficiency of measurements, it is 
impractical to use all these variables to form the state space for MLE estimation. 
It is 
more feasible to reconstruct the power system dynamics only with a small number of 
state variables. 

According to Takens' theorem \cite{takens1981detecting, packard1980geometry}, the dynamics of a nonlinear system can be reconstructed by the observations of a single state variable, and the reconstructed state vector can be expressed as
\begin{align}\label{eq:2}
\boldsymbol{\Theta}_t=\left[ {\theta_t}, \theta_{t-\tau}, \theta_{t-2\tau},\cdots,\theta_{t-2N\tau} \right]^\top,
\end{align}
where
$N$ is the dimension of the original system,
$\tau$ is the lag time,
and $\theta_t$, $\theta_{t-\tau}$, $\theta_{t-2\tau}$, $\cdots$, $\theta_{t-2N\tau}$ are the observations of the observed variable at the corresponding time steps.

In the reconstructed state space, the Euclidean distance in (\ref{eq:1}) becomes
\begin{align}\label{eq:3}
d(m(n),n,i) = ||\boldsymbol{\Theta}_{m(n)+i}-\boldsymbol{\Theta}_{n+i}||, \ i \in \{0,k\},
\end{align}
where 
$\boldsymbol{\Theta}_{n+i}$ and $\boldsymbol{\Theta}_{m(n)+i}$ are the observations on the reconstructed trajectory.

Moreover, another 
simplification 
can be made 
based on 
\cite{skokos2016chaos} and \cite{kantz1994robust}, 
in which it is shown that it is sufficient to only consider the first component of the reconstructed state vector to estimate MLE, 
because all components will grow exponentially at the rate of MLE.
Therefore, the $\boldsymbol{\Theta}_{n+i}$ and $\boldsymbol{\Theta}_{m(n)+i}$ in (\ref{eq:3}) can be replaced by their first components, and the distance in (\ref{eq:1}) can be calculated by
\begin{align}\label{eq:4}
d(m(n),n,i) = |{\theta}_{m(n)+i}-{\theta}_{n+i}|, \ i \in \{0,k\}. 
\end{align}

In this paper, the relative rotor angle of the severely disturbed generator pair (SDGP) is selected as the observed state variable, 
because these SDGPs 
are, in general, responsible for the system dynamics after considerable disturbances \cite{haque1989determination}.
An SDGP should be composed of a severely disturbed generator and the least disturbed generator 
in order to reflect the dynamics of the severely disturbed generators \cite{yin2011improved}.
In particular, the SDGPs can be identified as follows. 

\begin{enumerate}
	\vspace*{0.1cm}
	\item Obtain the 
	rotor speed of all generators at the fault clearing moment, $\omega_{t_c,n},\ n=1,2,\dots,N_G$, from PMU measurements, where $N_G$ is the number of generators.
	
	\vspace*{0.1cm} 
	\item Obtain the maximal absolute value of the rotor speed, i.e., $\omega_{t_c}^*=\max\limits_{n=1,2,\ldots,N_G}\left|\omega_{t_c,n}\right|$. 
	Define generator $g$ as one of the severely disturbed generators, if $\left|\omega_{t_c,g}\right|/\omega_{t_c}^*>\sigma$, where $\sigma$ is a predetermined threshold. In this paper $\sigma$ is chosen as 0.7, as in \cite{haque1995further}.
	
	\vspace*{0.1cm}
	\item Find the least disturbed generator with the minimal absolute value of the rotor speed.
	
	\vspace*{0.1cm}
	\item Form a SDGP by combing one of the severely disturbed generators and the least disturbed generator.
	Iterate over the severely disturbed generators and form all SDGPs.
\end{enumerate}

\subsection{RLS-Based MLE Estimation}

The MLE can be estimated by calculating the slope of the logarithmic distance curve in Phase II.
Considering the influences of measurement errors and nonlinear fluctuations, we adopt the least square algorithm 
to estimate the MLE.

From (\ref{eq:1}) and Fig. \ref{F2} it is seen 
that the MLE is estimated starting from time step $m(n)$.
From this time step, $k+1$ sequential logarithmic distances 
can be obtained as
\begin{align}\label{eq:4_5}
L(m(n)+i)&=\log\left(d\left(m(n),n,i\right)\right) \nonumber
\\
&=\log\left(|{\theta}_{m(n)+i}-{\theta}_{n+i}|\right), i=0,1,\dots,k. 
\end{align}

Then, the MLE estimation model can be expressed as
\begin{align}\label{eq:5}
L(m(n)+i)=\lambda_k\cdot (m(n)+i)\Delta t + C_k+\xi_k, \nonumber
\\
i=0,1,\dots,k,
\end{align}
where 
$\lambda_k$ is the MLE to be estimated,
$C_k$ is the constant term,
and $\xi_k$ is the residual term.

The solution of the least square estimation is
\begin{align}\label{eq:6}
\hat{\boldsymbol{E}}_k=
\left[\begin{array}{c}
\lambda_k\\
C_k
\end{array}\right]
=(\boldsymbol{X}_k^\top\boldsymbol{X}_k)^{-1} \boldsymbol{X}_k^\top \boldsymbol{Y}_k,
\end{align}
where
$\boldsymbol{X}_k$ is the coefficient matrix
and $\boldsymbol{Y}_k$ is the observation vector.
According to the aforementioned definitions, $\boldsymbol{X}_k$ and $\boldsymbol{Y}_k$ can be expressed as
\begin{small}
\begin{align}\label{eq:7}
&\boldsymbol{X}_k\!=\!
\left[\!\begin{array}{c c}
m(n)\Delta     t & 1\\
(m(n)+1)\Delta t & 1\\
\cdots          & \cdots\\
(m(n)+k)\Delta t & 1
\end{array}\!\right], \\ \nonumber \\ 
&\boldsymbol{Y}_k\!=\!
\left[\!\begin{array}{c}
L(m(n))\\
L(m(n)+1)\\
\cdots\\
L(m(n)+k)\\
\end{array}\!\right].
\end{align}
\end{small}
 
Moreover, to avoid the repeated calculation of the inverse matrix in (\ref{eq:6}), a recursive estimation algorithm is applied \cite{young2011recursive}, which can be formulated as
\begin{align}\label{eq:8}
\nonumber&\hat{\boldsymbol{E}}_{k+1}= \hat{\boldsymbol{E}}_{k}+\boldsymbol{G}_{k+1}\left[y_{k+1}-\boldsymbol{x}_{k+1}^\top\hat{\boldsymbol{E}}_k\right],\\
\nonumber&\boldsymbol{G}_{k+1}=\frac{\boldsymbol{P}_k\boldsymbol{x}_{k+1}}{1+\boldsymbol{x}_{k+1}^\top\boldsymbol{P}_k\boldsymbol{x}_{k+1}},\\
&\boldsymbol{P}_{k+1}=\boldsymbol{P}_k-\boldsymbol{G}_{k+1}\boldsymbol{x}_{k+1}^\top\boldsymbol{P}_k,
\end{align}
where
$\boldsymbol{x}_{k+1}$ is $\left[(m(n)+k+1)\Delta t \ \ 1\right]^\top$,
$y_{k+1}$ is the new observation $L(m(n)+k+1)$,
$\hat{\boldsymbol{E}}_{k}$ and $\hat{\boldsymbol{E}}_{k+1}$ are the estimation results before and after obtaining the new observation,
$\boldsymbol{P}_k$ and $\boldsymbol{P}_{k+1}$ are the covariance matrices,
and $\boldsymbol{G}_{k+1}$ is the gain vector.

The algorithm includes the following three steps:
\begin{enumerate}
	\item With the first 2 groups of data ($k=1$ in (\ref{eq:4_5})), set the initial values of $\hat{\boldsymbol{E}}_{1}$ and $\boldsymbol{P}_1$ to be $(\boldsymbol{X}_1^\top\boldsymbol{X}_1)^{-1} \boldsymbol{X}_1^\top \boldsymbol{Y}_1$ and $(\boldsymbol{X}_1^\top\boldsymbol{X}_1)^{-1}$, respectively. 
	\vspace*{0.1cm}
	\item Obtain a new observation of the logarithmic distance between the trajectories, and then sequentially calculate $\boldsymbol{G}_{k+1}$, $\hat{\boldsymbol{E}}_{k+1}$, and  $\boldsymbol{P}_{k+1}$ according to (\ref{eq:8}).
	\vspace*{0.1cm}
	\item Set $k=k+1$ and return to step 2.
\end{enumerate}


\subsection{Parameter Setting}

Because the Theiler window $w$ and the MLE estimation initial time step $m(n)$ determine the shape of the estimated MLE curve, 
they are crucial for a quick 
and reliable rotor angle stability assessment.
Here, we discuss how to choose these parameters according to the rotor angle swing features.

\vspace{0.2cm}
\noindent \textit{(1) Theiler window selection}
\vspace{0.2cm}

The Theiler window $w$ determines the temporal separation between the initial points $\theta_{0}$ and $\theta_{m(0)}$.
It should be large enough to ensure that $\theta_{0}$ and $\theta_{m(0)}$ are the initial points of different trajectories.
However, too large $w$ will cause unnecessary 
waiting time and 
delay the stability assessment. 

\begin{figure}[!b]
	\centering
	\includegraphics[width=3.4in]{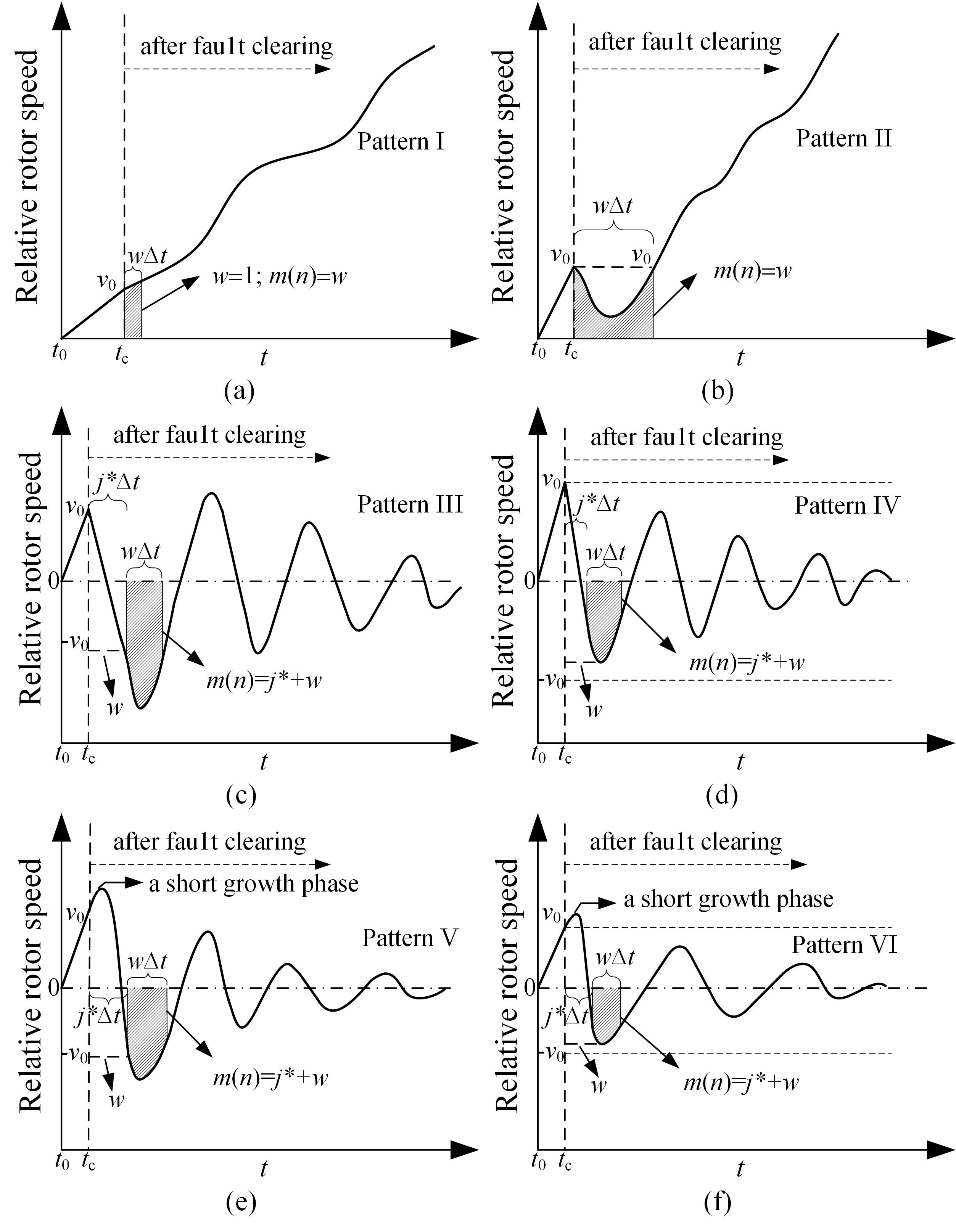}
	\caption{Relative rotor speed curves of different patterns.}
	\label{F3}
\end{figure}

The post-fault rotor angles of the SDGPs have significant swing patterns \cite{spalding1977coherency,takahashi1988fast,karady2002line}.
According to the 
features of the relative rotor speed of the SDGPs, six distinct swing patterns can be identified as shown in Fig. \ref{F3}.
Different $w$ 
are chosen for different patterns as follows.

\begin{itemize}
\item \textbf{Pattern I}:
In this pattern, the relative rotor speed 
increases after fault clearing.
No decelerating area exists for the SDGP and the system will lose stability during the first swing. 
Since the original and the neighboring trajectories separate rapidly, $w$ is set to 1 (the smallest positive integer) to minimize the estimation waiting time.

\vspace*{0.2cm}
\item \textbf{Pattern II}:
In this pattern, the relative rotor speed decreases after fault clearing.
However, since the decelerating area is relatively small, the relative rotor speed increases again after a short time period, 
and the increasing trend continues until the system loses stability.
The key feature of this pattern is that the relative rotor speed at the fault clearing moment, denoted by $v_0$ in Fig. \ref{F3}, appears again after the initial decrease.
In order to achieve obvious separation between the trajectories, $w$ is set to be the time step lags of the reappearance of $v_0$.

\vspace*{0.2cm}
\item \textbf{Pattern III}:
In this pattern, the relative rotor speed first decreases to $-v_0$, and then exhibits periodic oscillations.
The decelerating area is large enough to reduce the relative rotor speed to zero, and the system stability depends on the damping characteristics of the post-fault equilibrium point.
In this case, $w$ is set to be the time step lags of the first appearance of $-v_0$.

\vspace*{0.2cm}
\item \textbf{Pattern IV}:
In this pattern, the relative rotor speed first decreases to some value greater than $-v_0$, and then oscillates periodically.
The key feature of this pattern is that the relative rotor speed $v_0$ and $-v_0$ cannot be observed after fault clearing.
This pattern is a special case of Pattern III, and $w$ is set to be the time step lags of the first appearance of the local minimal relative rotor speed after fault clearing.

\vspace*{0.2cm}
\item \textbf{Pattern V}:
This pattern is similar to Pattern III, and usually appears after very quick fault clearing.
The key feature of this pattern is that the relative rotor speed shows decelerated growth 
immediately after fault clearing, and the relative rotor speed $-v_0$ can be observed after that.
The $w$ is set in the same way as in Pattern III. 

\vspace*{0.2cm}
\item \textbf{Pattern VI}:
This pattern is a special case of Pattern V.
The key feature of this pattern is that the relative rotor speed shows decelerated growth immediately after the fault clearing, and the relative rotor speed $-v_0$ cannot be observed during the oscillations.
The $w$ is set in the same way as in Pattern IV.
\end{itemize}

\vspace{0.2cm}
\noindent \textit{(2) MLE estimation initial time step selection}
\vspace{0.2cm}
 
The MLE estimation initial time step $m(n)$ should ensure that the slope estimation is performed for Phase II of the logarithmic distance growth. 
In this phase, the original and neighboring trajectories have been sufficiently separated, and thus the logarithmic distance curve has shown clear development trend \cite{skokos2016chaos}.

In fact, the distance between the rotor angle trajectories has a close relationship with the relative rotor speed of the SDGP.
According to the definition of the Theiler window, 
the distance between the corresponding points of the original and the neighboring trajectories at any time step $j$ can be reformulated as
\begin{align}\label{eq:9}
d_j &\!=\! d(m(0),0,j) \nonumber\!=\!\left|\theta_{m(0)+j}-\theta_{0+j}\right|
\\
& \! \approx\! \left|\!\left(\!{\theta}_{0+j}\!+\!\sum_{t=j+1}^{j+w}v_t\!\cdot\!\Delta t\!\right)\!\!-\!{\theta}_{0+j}\!\right|
\!=\! \left|\sum_{t=j+1}^{j+w}v_t\!\cdot\!\Delta t\right|,
\end{align}
where $v_t$ is the relative rotor speed at the relevant time step.

It is revealed in (\ref{eq:9}) that the distance $d_j$ is equal to the area enclosed by the time axis and the relative rotor speed curve within the Theiler window, 
as indicated by the shaded regions in Fig. 3.
Therefore, 
$m(n)$ can be determined 
as follows.

\begin{itemize}
\vspace*{0.2cm}
\item \textbf{Patterns I--II}:
With the selected $w$, it is seen in Fig. \ref{F3} that the distance $d_j$ of these patterns monotonically increases after fault clearing, which indicates that the logarithmic distance growth is in Phase II as soon as the fault is cleared.
Therefore, in these two patterns, $m(n)$ is set to be $w$ to minimize the estimation waiting time.

\vspace*{0.2cm}
\item \textbf{Patterns III--VI}: 
For these patterns 
the distance will exhibit periodic fluctuations after fault clearing.
To ensure the trajectories have been sufficiently separated and to catch the main trend of the fluctuations, 
$m(n)$ is set to be $w+j^*$, where $j^*$ is the time step when 
$d_{j}$ reaches its first local maximum after fault clearing.
\end{itemize}

\subsection{Rotor Angle Stability Assessment Criteria}

When the MLE curve of a SDGP is estimated, the following criteria can be used to 
determine the stability of the SDGP.

\begin{itemize}
\vspace*{0.1cm}
\item \textbf{Criterion I}:
If the MLE of the SDGP increases at the beginning, the SDGP is unstable.

\vspace*{0.1cm}
\item \textbf{Criterion II}:
If the MLE decreases at the beginning, it will have oscillations. 
If the first peak point of the oscillation is positive, the SDGP is unstable.

\vspace*{0.1cm}
\item \textbf{Criterion III}:
If the MLE decreases at the beginning and the first peak point of the oscillation is negative, the SDGP is stable.

\end{itemize}

\begin{figure}[!b]
	\centering
	\includegraphics[width=2.1in]{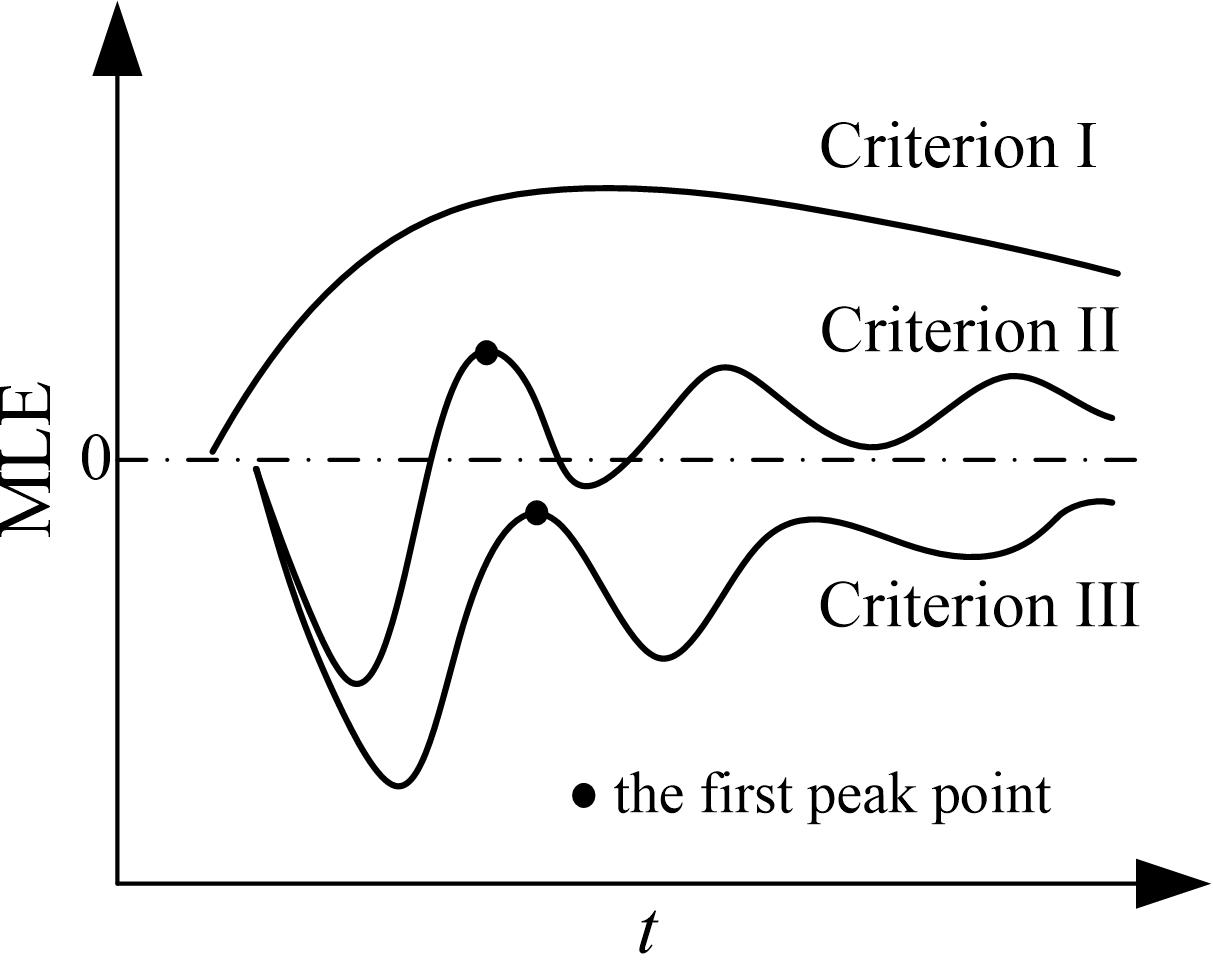}
	\caption{Typical MLE curves for Criterion I--III.}
	\label{F4}
\end{figure}

Typical MLE curves corresponding to these criteria are shown in Fig. \ref{F4}. 
If the condition in Criterion I is satisfied, 
Pattern I or II in Fig. \ref{F3} will happen, 
for which the distance will increase immediately after the MLE estimation initial time step and the trend will last until the SDGP loses stability.

\begin{figure}[!htb]
	\centering
	\includegraphics[width=3.4in]{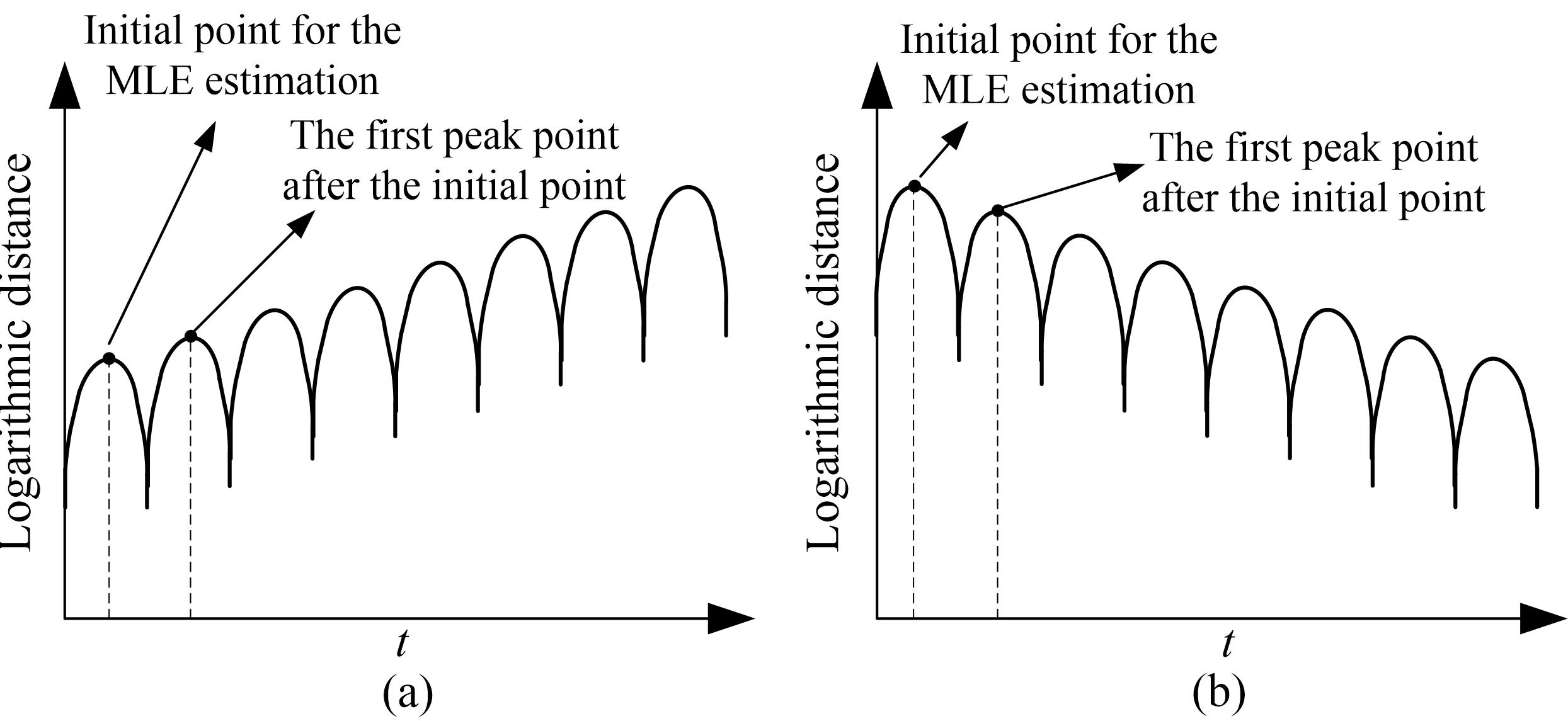}
	\caption{The logarithmic distance curves for Pattern III--VI.}
	\label{F5}
\end{figure}

Criteria II and III correspond to Patterns III--VI.
In these patterns, the relative rotor angle will oscillate after fault clearing.
However, by using the selected parameters, the logarithmic distance will exhibit significant development trend as illustrated in Fig. \ref{F5}. 
If the relative rotor speed has a 
undamped oscillation, the SDGP is unstable and 
the logarithmic distance curve will fluctuate periodically as shown in Fig. \ref{F5}(a).
Since the MLE is the average slope of the logarithmic distance curve from the MLE estimation initial time step, 
it will first decrease and then increase to a positive peak value, as in Fig. \ref{F4}. 
By contrast, if the relative rotor speed has a damped oscillation, the SDGP is stable and the logarithmic distance curve will look like Fig. \ref{F5}(b). 
The MLE will first decrease and then increase to a negative peak value, 
as in Fig. \ref{F4}.

Because SDGPs are responsible for the system dynamics after disturbances \cite{haque1989determination}, 
we finally have the following criterion on the angle stability of the system.

\begin{itemize}
\vspace*{0.1cm}
\item \textbf{Criterion IV}:
If all SDGPs are stable, the system is stable; 
otherwise, the system is unstable.   
\end{itemize}

\subsection{Rotor Angle Stability Assessment Procedure}

\begin{figure}[!htb]
	\centering
	\includegraphics[width=3.4in]{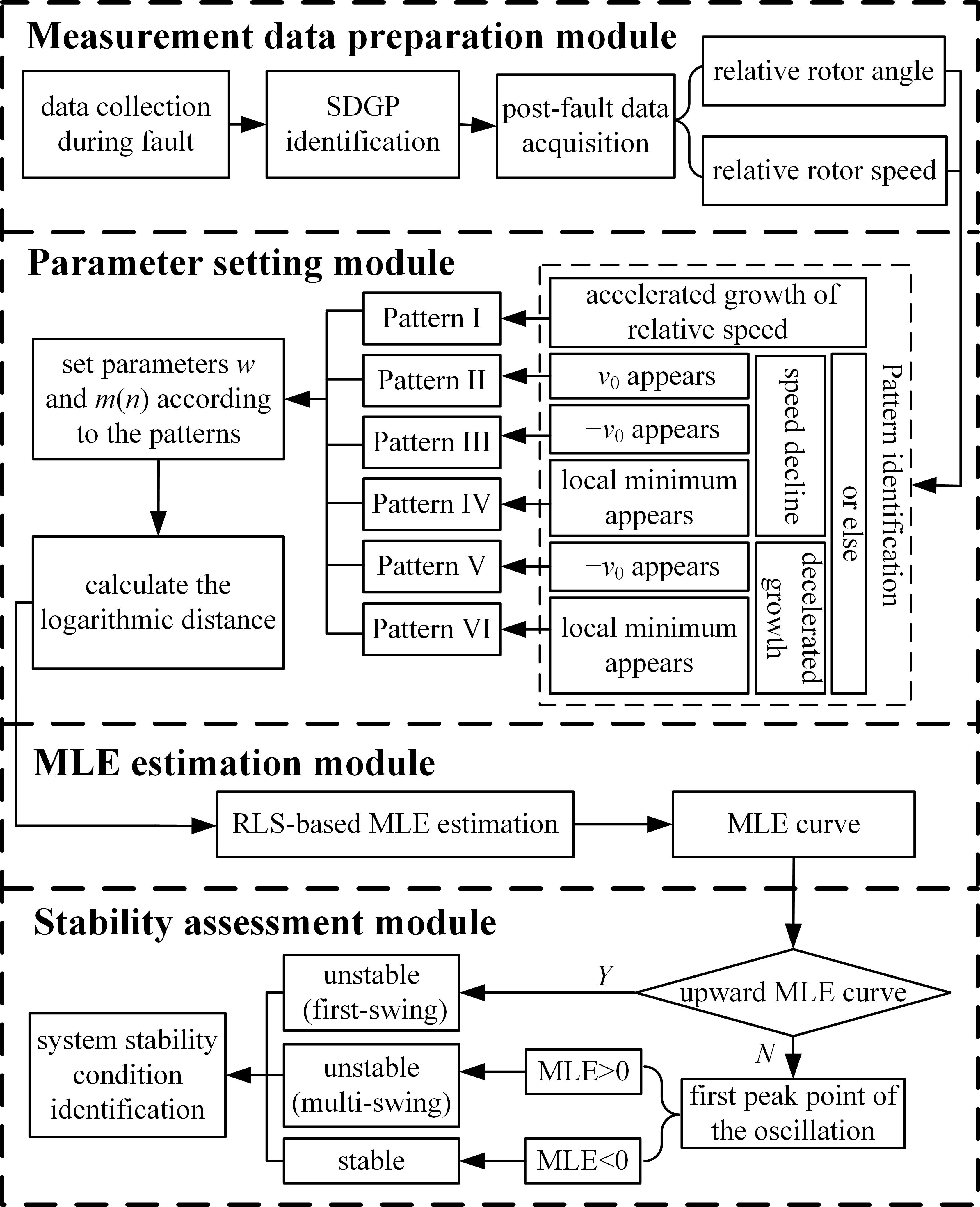}
	\caption{Flow chart of the assessment procedure.}
	\label{F6}
\end{figure}

The proposed online rotor angle stability assessment procedure 
is shown in Fig. \ref{F6},
which includes 
the measurement data preparation module, the parameter setting module, the MLE estimation module, and the stability assessment module. 
Specifically,

\begin{enumerate}
\item When a fault is detected, the measurement data preparation module will be immediately activated to identify the SDGPs and collect the corresponding relative rotor angle and rotor speed measurements in real time.
\vspace*{0.1cm}
\item The relative rotor speed variation patterns are identified online according to the rules described in Section III-C, and the parameters are assigned accordingly.
\vspace*{0.1cm}
\item The MLE sequences can be calculated 
by using the RLS-based algorithm in Section III-B.
\vspace*{0.1cm}
\item Finally the stability condition can be assessed by using the criteria provided in Section III-D according to the features of the estimated MLE curves.

\end{enumerate}

\section{Case Studies}

The proposed approach is tested on the New-England 39-bus system and the NPCC 140-bus system. Simulations are performed with Power System Analysis Toolbox (PSAT) \cite{milano2005open} in MATLAB.

\subsection{New-England 39-Bus System}

The New-England 39-bus system has 10 generators and 46 branches. The parameters can be found in \cite{pai1989energy}.
Unless otherwise specified, all generators in the tests are described by the fourth-order transient model with Type I turbine governor (TG), Type II automatic voltage regulator (AVR), and Type II power system stabilizer (PSS) (see PSAT documentation).
All loads are described by the ZIP model and the ratios of the constant impedance, constant current, and constant power loads are $0.4$, $0.5$, and $0.1$, respectively. 
The sampling rate of the PMU measurements used for MLE estimation is 120 samples/s
\cite{dasgupta2015pmu}.

To verify the effectiveness of the proposed approach, a three-phase-to-ground fault is applied at bus 2 at $t=1$ s, and the fault is cleared by opening line $2$--$3$ at $t_c=1.243$ s and $t_c=1.244$ s for Scenarios I and II, respectively.
According to the time-domain simulation, the system is stable under Scenario I and unstable under Scenario II.

\begin{figure}[!b]
	\centering
	\includegraphics[width=3.3in]{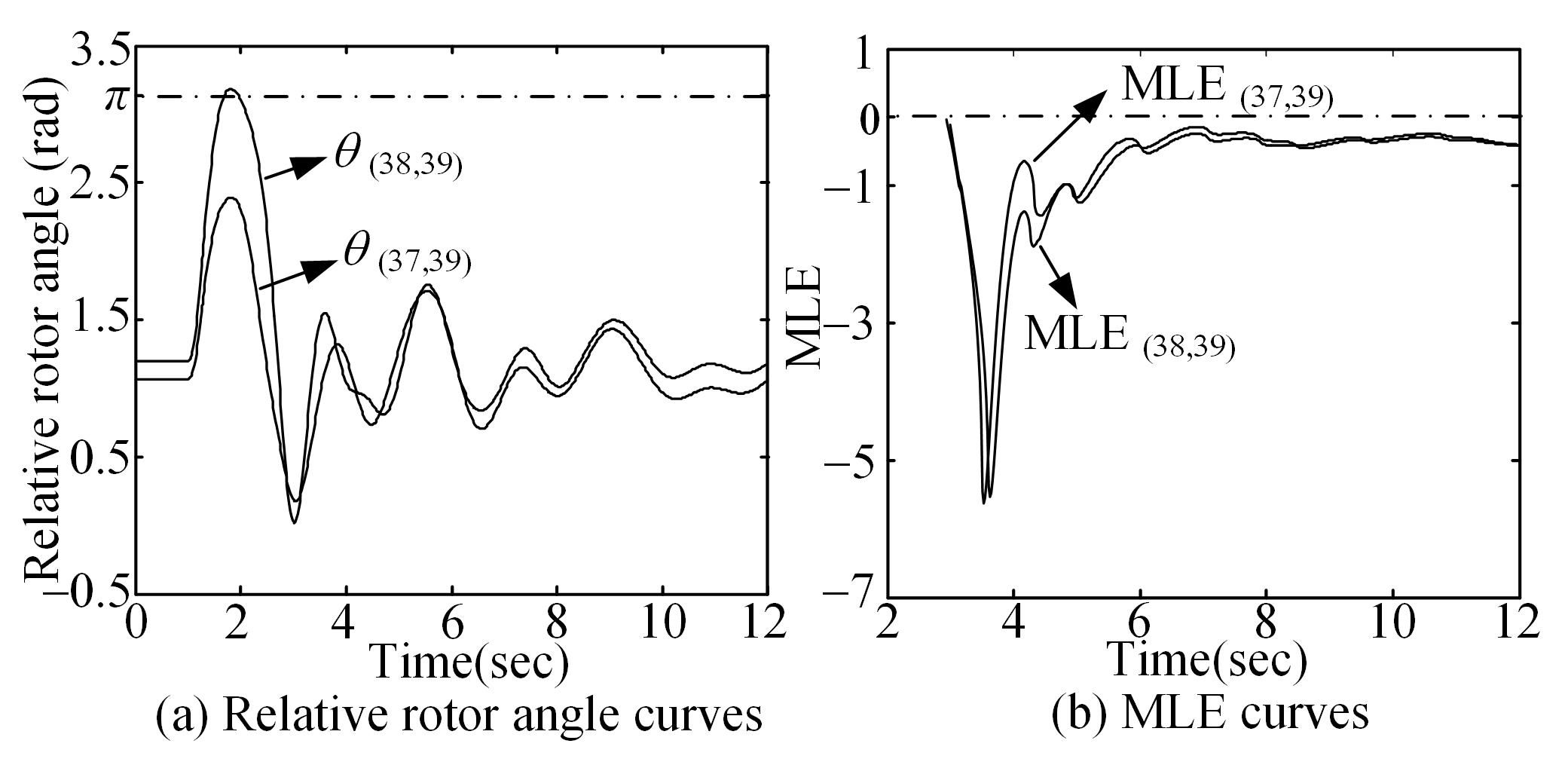}
	\caption{Simulation results of Scenario I.}
	\label{F7}
\end{figure}

\begin{figure}[!htb]
	\centering
	\includegraphics[width=3.3in]{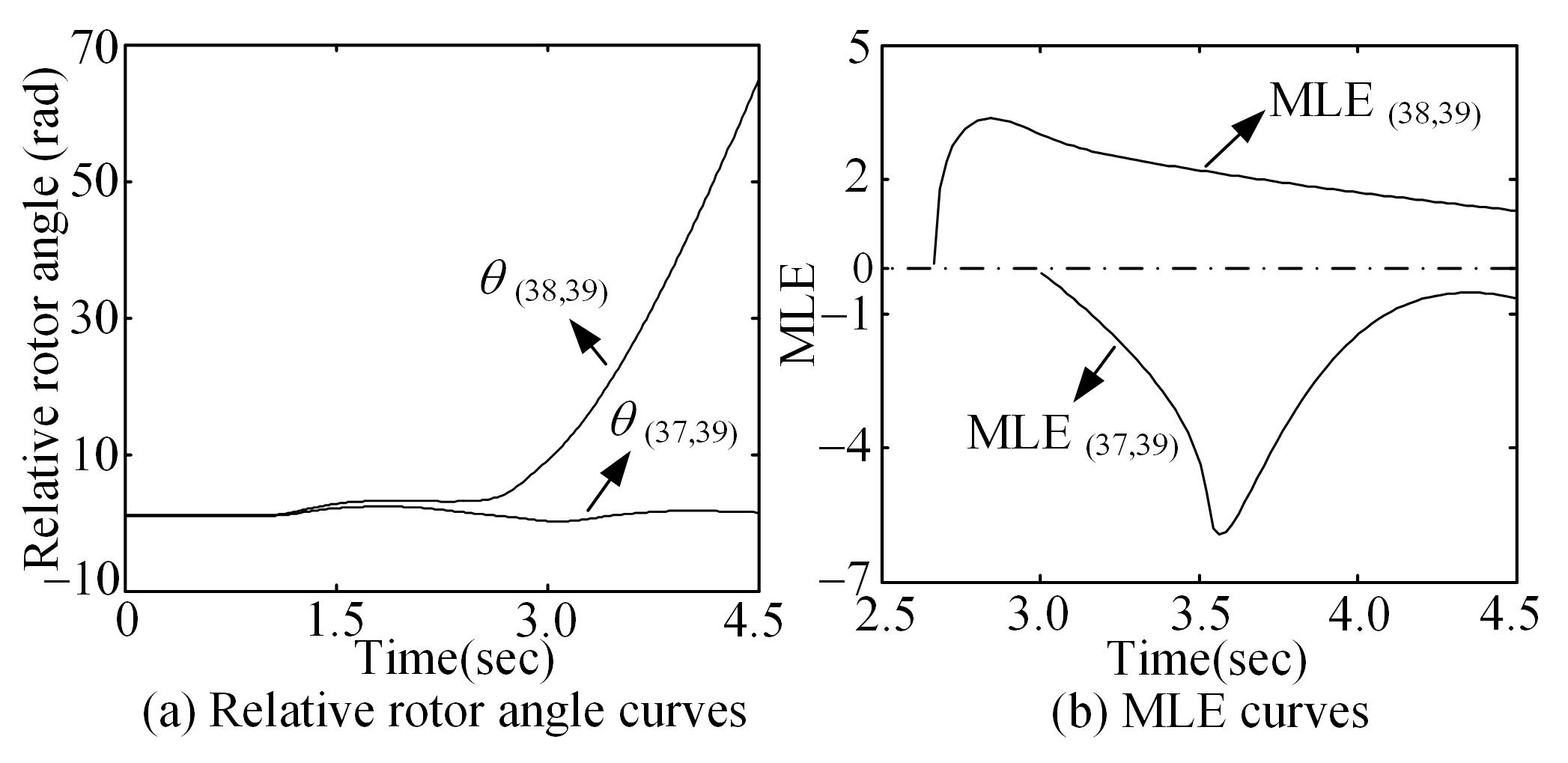}
	\caption{Simulation results of Scenario II.}
	\label{F8}
\end{figure}

By using the method in Section III-A, the generator pairs $38$--$39$ and $37$--$39$ are identified as the SDGPs for both scenarios.
Figs. \ref{F7}--\ref{F8} show the relative rotor angles and the estimated MLEs of the SDGPs.

From Fig. \ref{F7}(a), it is seen that the relative rotor angles of the SDGPs tend to be stable after a long period of oscillations.
In Fig. \ref{F7}(b), the MLEs of the SDGPs decrease immediately after fault clearing, and their first peak points of the oscillations are both less than 0.
Therefore, according to Criteria III and IV, 
the system is stable under this scenario.

By contrast, under Scenario II the MLE of the generator pair $38$--$39$ increases immediately after fault clearing, as shown in Fig. \ref{F8}(b).
Therefore, according to Criteria I and IV, the system is unstable under this scenario.

For the two scenarios, the system stability can be assessed, respectively, within $2.82$ s and $1.40$ s after fault clearing, which demonstrates that the proposed approach can provide early detection of instability.
In fact, according to the proposed approach, the first-swing instability can be identified very quickly, because the instability can be detected at the beginning of the MLE curve.
On the other hand, for the multi-swing stability or instability, the assessment time is a little longer since the approach needs to check the first peak point of the MLE curve after fault clearing.

In industrial applications, a predetermined relative rotor angle value, i.e., $\pi$ rad, is usually set as the threshold for determining rotor angle stability \cite{xu2012reliable}.
Although this pragmatic criterion is easy to execute, the 
relative rotor angle corresponding to rotor angle instability 
may change significantly with the change of topologies, parameters, and operating conditions.
For instance, under Scenario I, the relative rotor angle between generators $38$ and $39$ can reach up to $3.207$ rad while the system is still stable.

It should also be noted that there are foundational differences between the proposed approach and that in \cite{dasgupta2015pmu}.
In the proposed approach, the system stability can be explicitly 
determined 
at latest when the 
first peak point of the MLE curve is observed. 
By contrast, 
for the approach in \cite{dasgupta2015pmu}, the MLE curve must be observed for a long period to ensure the sign of MLE.
Unfortunately, it is actually difficult for the approach in \cite{dasgupta2015pmu} to predetermine the observing window size to provide reliable and timely assessment results (for instance, see Figs. 1--2 in \cite{dasgupta2015pmu}). 
From this perspective, the proposed approach is more time efficient and more reliable. 

In order to further verify the accuracy of the proposed approach, extensive tests are performed on the New-England 39-bus system.
Specifically, a three-phase-to-ground fault is created for each bus (except the generator buses) at $t=1$ s, and is cleared at $1.08$ s, $1.16$ s, $1.24$ s, and $1.32$ s, respectively.
According to the test results, the proposed approach successfully determines the rotor angle stability in all $224$ tests.
The occurrence frequencies of the fault patterns 
and the success rate of the proposed stability assessment approach
are summarized in Table \ref{Table_1}.

\begin{table}[!htb]
	\renewcommand{\arraystretch}{1.2}
	\caption{Summary of the Tests on the New-England 39-Bus System}
	\vspace*{-6pt}
	\label{Table_1}
	\centering
	\begin{tabular}{c c c c c c c}
		\hline
		\multirow{2}{*}{$t_c /$s}  & \multicolumn{6}{c}{Occurrence Times of the Swing Patterns}\\
		\cline{2-7}
        &I & II & III & IV & V & VI \\
        \hline
	     1.08& N/A & N/A & 4& 43& 6& 3\\
		 1.16& N/A & 6& 10& 36& 3& 1\\
		 1.24& 4& 13& 9& 27& 2& 1\\
		 1.32& 38& 12& 2& 4&N/A &N/A \\
Success Rate &100\%&100\%&100\%&100\%&100\%&100\%\\
 \hline
	\end{tabular}
\end{table}

It is seen in Table \ref{Table_1} that Patterns I and II those correspond to first-swing instability usually occur with longer fault clearing time.
In contrast, Patterns V and IV usually occur with the shorter fault clearing time, and the system will have a good chance to maintain stability in these patterns.
Among the tests, the first-swing instability identification time ranges from $1.2$ s to $1.5$ s, and the multi-swing stability assessment time ranges from $2.2$ s to $2.5$ s,
which further confirms the efficient of the proposed approach.

\begin{figure}[!htb]
	\centering
	\includegraphics[width=3.3in]{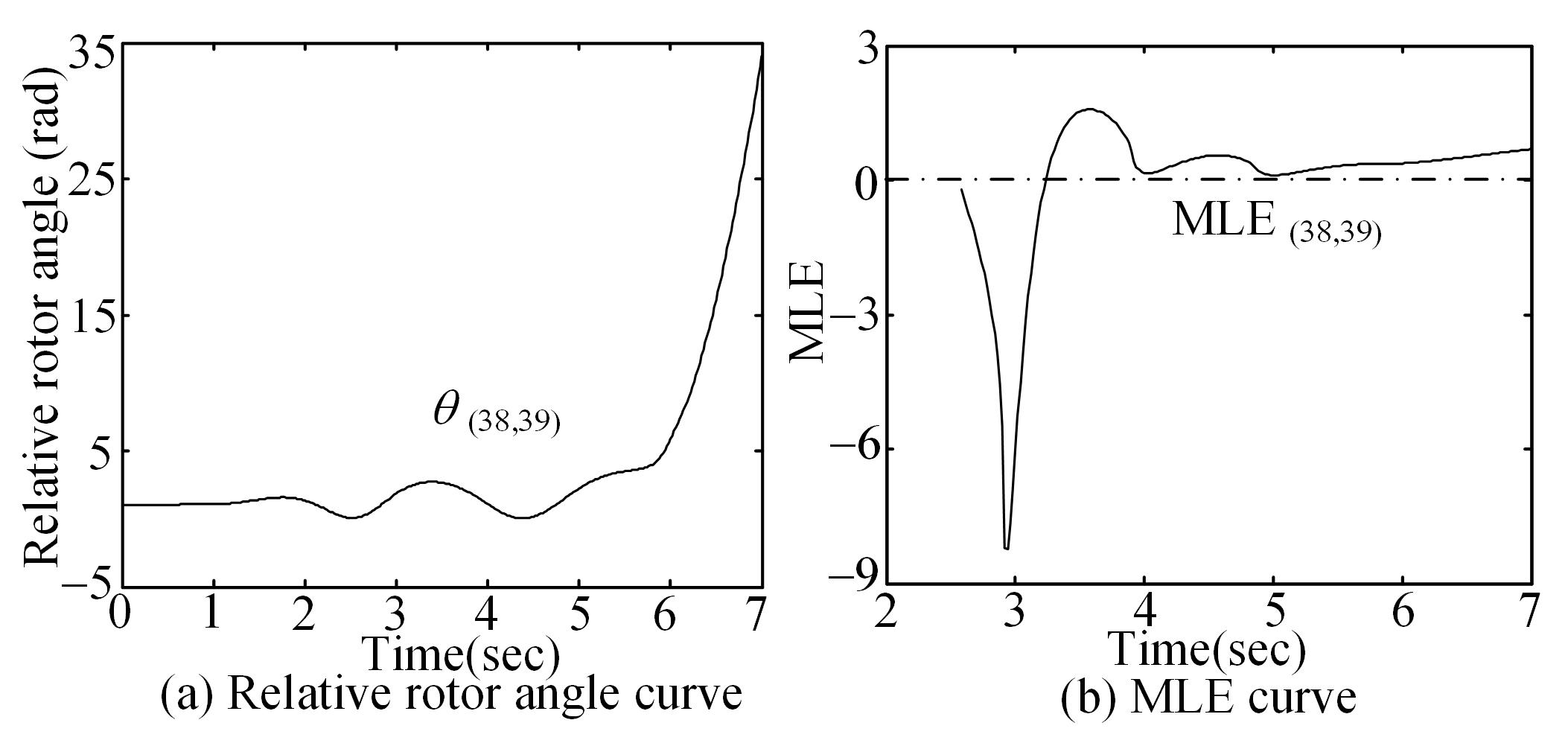}
	\caption{Simulation results of the multi-swing unstable case.}
	\label{F9}
\end{figure}

Because the system has sufficient damping to ensure the stability of the post-fault equilibrium points, 
there are no multi-swing instability cases. 
In order to generate a multi-swing instability case, 
all of the PSSs are removed and the parameters of the AVRs are tuned. 
In the modified system, a three-phase-to-ground fault is applied at bus 28 at $t=1$ s, and is cleared by opening line $27$--$28$ at $t_c=1.12$ s.
In this test, only generator pair $38$--$39$ is identified as the SDGP, whose relative rotor angle curve and MLE curve are shown in Fig. \ref{F9}.

In Fig. \ref{F9}(b), the MLE decreases immediately after fault clearing, and the following peak MLE is positive.
According to Criteria II and IV, the MLE variation pattern indicates that the system is multi-swing unstable, which is verified by the relative rotor angle curve.
In this case, the stability assessment time is $2.36$ s, which is much shorter than the time required for directly looking at the relative rotor angle curve. 

\subsection{NPCC 140-Bus System}

The proposed approach is also tested on the NPCC 140-bus system 
\cite{chow1995inertial}.
The settings are the same as those for the New-England 39-bus system.

\begin{figure}[!htb]
	\centering
	\includegraphics[width=3.3in]{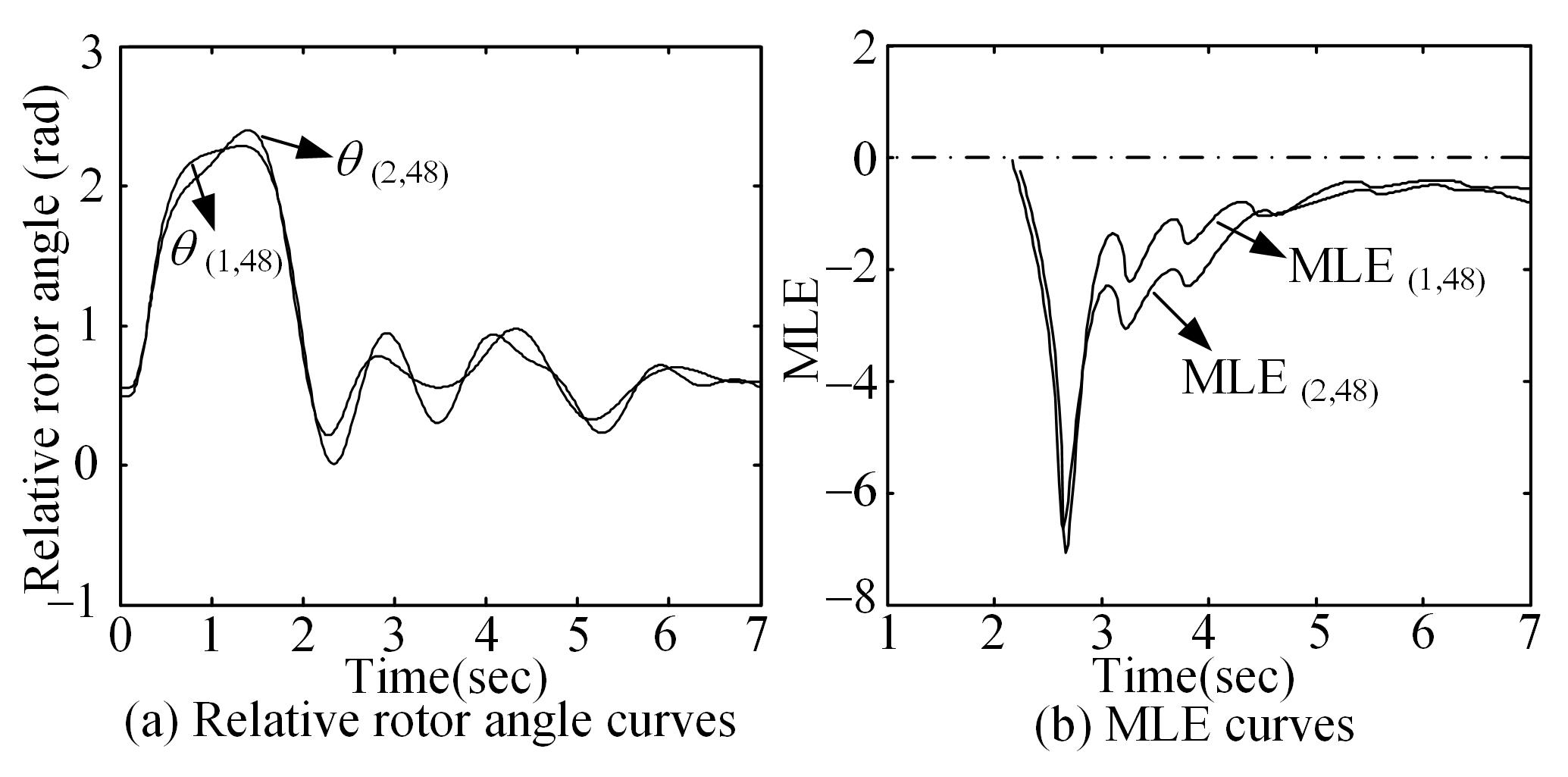}
	\caption{Simulation results of Scenario III.}
	\label{F10}
\end{figure}

\begin{figure}[!htb]
	\centering
	\includegraphics[width=3.3in]{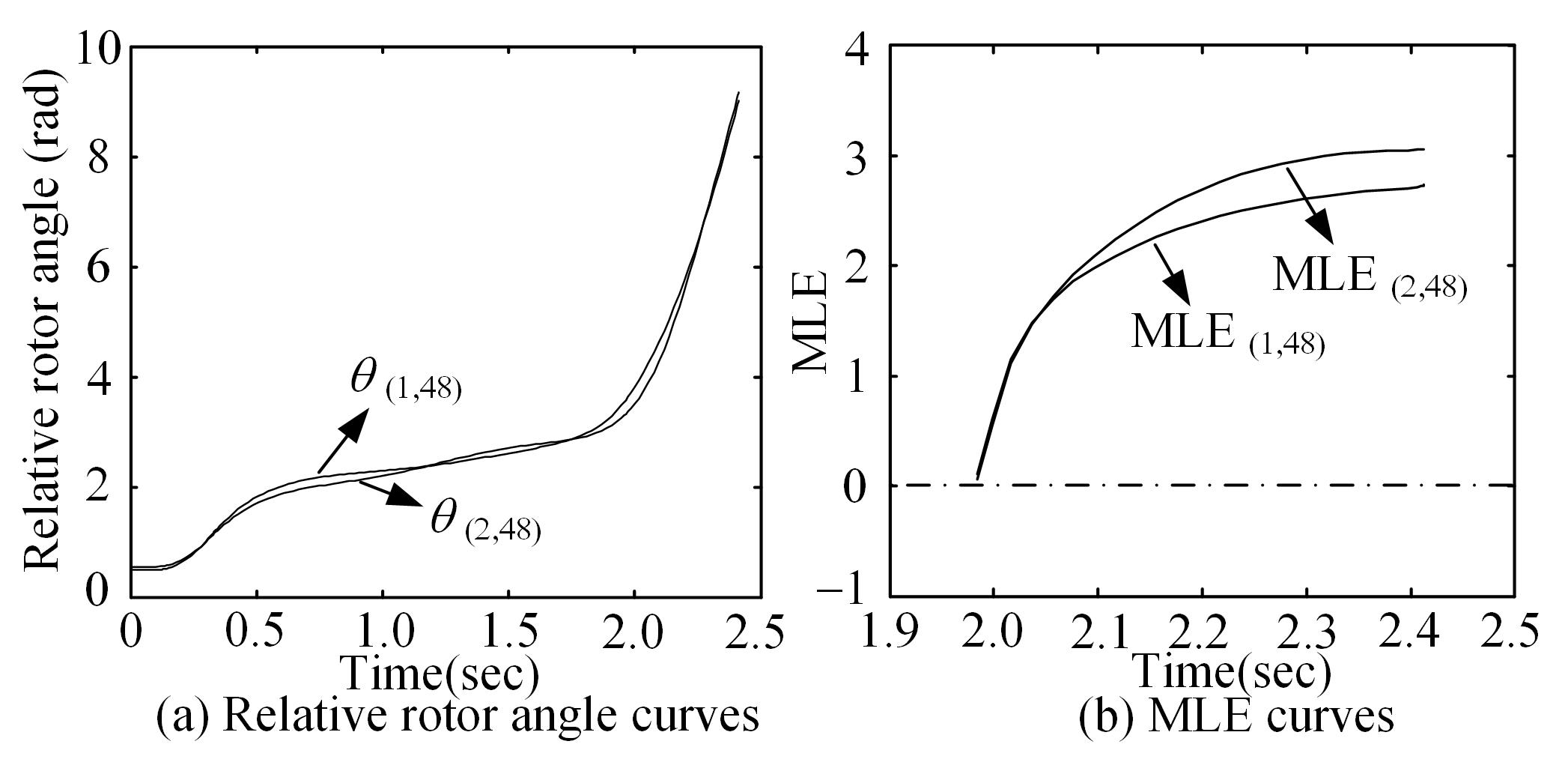}
	\caption{Simulation results of Scenario IV.}
	\label{F11}
\end{figure}

A three-phase-to-ground fault is first applied at bus $35$ at $t=0.1$ s, and is cleared by opening line $34$--$35$ at $t_c=0.307$ s (Scenario III, stable) and $t_c=0.308$ s (Scenario IV, unstable), respectively.
Generator pairs $1$--$48$ and $2$--$48$ are identified as the SDGPs for both cases.
The relative rotor angle curves and MLE curves under these two scenarios are shown in Figs. \ref{F10}--\ref{F11}, respectively.

In Fig. \ref{F10}, it is seen that the curves of Scenario III are similar to those of Scenario I.
In this case, the system is determined as stable because of the same reason as Scenario I.
As shown in Fig. \ref{F11}, under Scenario IV the MLE curves of both SDGPs exhibit unstable features, which indicates that both generator pairs will lose stability
and thus the system is unstable. 

Extensive tests are also executed on the NPCC 140-bus system.
A three-phase-to-ground fault is applied at each bus (except the generator buses) at $t=0.1$ s, and is cleared at $0.18$ s, $0.26$ s, $0.32$ s and $0.40$ s, respectively.
Table \ref{Table_2} lists the occurrence frequencies of the fault patterns and the success rate of the proposed stability assessment approach during the tests.

\begin{table}[!htb]
	\renewcommand{\arraystretch}{1.2}
	\caption{Summary of the Tests on the NPCC 140-Bus System}
	\vspace*{-6pt}
	\label{Table_2}
	\centering
	\begin{tabular}{c c c c c c c}
		\hline
		\multirow{2}{*}{$t_c /$s}  & \multicolumn{6}{c}{Occurrence Times of the Swing Patterns}\\
		\cline{2-7}
		&I & II & III & IV & V & VI \\
		\hline
		0.18& N/A & N/A & 6  & 166 & 2   & 5   \\
		0.26& N/A & 10  & 26 & 143 & N/A & N/A \\
		0.32& 8   & 57  & 38 & 76  & N/A & N/A \\
		0.40 & 60  & 12  & 49 & 58  & N/A & N/A \\
		Success Rate& 100\%&100\%&100\%&100\%&100\%&100\% \\
		\hline
	\end{tabular}
\end{table}

According to the test results, the proposed approach can accurately 
determine system stability in all $716$ tests.
The first-swing instability identification time ranges from $1.1$ s to $1.7$ s, and the multi-swing stability assessment time ranges from $1.8$ s to $2.4$ s.

\section{Conclusion}

In this paper, a model-free approach for online rotor angle stability assessment is proposed based on MLE.
By using the proposed MLE estimation algorithm, parameter setting rules and the stability criteria, the approach can online identify the system stability condition with PMU measurements.
The approach does not need a predetermined observing window to identify the sign of the MLE, and can provide reliable and timely assessment results by analyzing the features of the estimated MLE curve.
To verify the performance of the proposed approach, extensive tests are performed on the New-England 39-bus system and the NPCC 140-bus system.
The proposed approach can successfully determine the system stability conditions in all 945 tests.
Moreover, among all the tests, the first-swing stability can be assessed within $1.7$ s and the multi-swing stability can be assessed within $2.5$ s.


%

%
%
%
%
%

\ifCLASSOPTIONcaptionsoff
  \newpage
\fi



\bibliographystyle{IEEEtran}
\bibliography{Reference}
\end{document}